\documentclass[aps,pre,twocolumn,groupedaddress,showpacs,floatfix]{revtex4}
\usepackage{amsmath}
\usepackage{amsfonts}
\usepackage{amssymb}
\usepackage{graphicx}
\usepackage{subfigure} 
\usepackage{bm} 
\begin{document}

\title{Pulse Velocity in a Granular Chain}
\author{Alexandre Rosas and Katja Lindenberg}
\affiliation{
Department of Chemistry and Biochemistry,
and Institute for Nonlinear Science
University of California San Diego
La Jolla, CA 92093-0340}
\date{\today}

\begin{abstract}
We discuss the applicability of two very different analytic
approaches to the study of pulse propagation in a chain of
particles interacting via a Hertz potential, namely, a
continuum model and a binary collision approximation. While both methods
capture some qualitative features equally well, the
first is quantitatively good for softer potentials and the latter is
better for harder potentials.
\end{abstract}
\maketitle

The physics of a chain of particles interacting via
a granular potential, i.e. a  potential that is repulsive under
loading and zero otherwise, remains a challenge  
despite a great deal of recent
work on the subject~\cite{mackay,senmanciu,sinkovits,rogers,coste,sen98,
naughton,hinch,hascoet, hascoet2,sen-book,hong,sen-pasi,wu,nakagawa}.
Theoretical studies of pulse dynamics in frictionless chains have
relied primarily on numerical solution of the equations of
motion~\cite{sinkovits,sen98,hinch,hascoet,hascoet2,sen-pasi,hascoet3}.
Analytic work has relied on two rather different approximations
with very little direct comparison between them.  One approach is 
based on continuum approximations to the equations
equations of motion followed by exact or approximate
solutions of these approximate
equations~\cite{nesterenko,hinch,hascoet3}.  This approach is expected
to give useful results when the pulse is not too narrow, i.e., when the
velocity distribution of the grains in the chain at any instant of
time is rather
smooth.  The other approach is based on 
phenomenology about properties of pairwise
(or at times three-body) collisions together with the
assumption that pulses are sufficiently narrow to involve only two or
three grains at any one time~\cite{wu,mohan,ourtwo}.  Among the
interesting quantities 
one aims to calculate with these approaches is the pulse velocity.  In
turn, successful calculation of the pulse velocity requires a good
understanding of the pulse width.

The standard generic model potential between monodisperse elastic
granules that repel upon overlap according to the Hertz law is given
by~\cite{hertz,landau}
\begin{equation}
\begin{array}{l l l}
V(\delta_{k,k+1})&=\frac{a}{n}|\delta|^{n}_{k,k+1}, \qquad &\delta\leq 0,\\ \\
V(\delta_{k,k+1})&= 0, \qquad &\delta >0.
\end{array}
\label{eq:hertz}
\end{equation}
Here
\begin{equation}
\delta_{k,k+1} \equiv y_k - y_{k+1},
\end{equation}
$ a$ is a constant that depends 
on the Young's modulus and Poisson's ratio, and
$y_k$ is the displacement of granule $k$ from its equilibrium position.
The exponent $n$ is $5/2$ for spheres, it is $2$ for cylinders, and
in general depends on geometry.

The displacement of the $k$-th granule ($k=1,2, \ldots, L$) in the chain
from its equilibrium position in a frictionless medium
is thus governed by the equation of motion
\begin{eqnarray}
m \frac{\mathrm{d}^2{y}_k}{\mathrm{d} \tau^2} &=& - a (y_k - y_{k+1})^{n-1}
\theta (y_k - y_{k+1})\notag\\
&&+ a (y_{k-1} - y_{k})^{n-1}
\theta (y_{k-1} - y_{k}).
\label{eq:motion}
\end{eqnarray}
Here $\theta(y)$ is the Heavyside function, $\theta(y)=1$ for $y>0$, 
$\theta(y)=0$ for $y<0$, and $\theta(0)=1/2$. 
It ensures that the particles interact only
when
in contact. Note that for a finite open chain the first term
on the right hand side of this equation
is absent for the last granule and the second term is absent for
the first.

Initially the granules are placed along a line so that they
just touch their neighbors in their equilibrium positions (no
precompression), and
all but the leftmost particle are at rest. The initial velocity of
the leftmost particle is $v_0$ (the impulse). 
In terms of the rescaled variables 
\begin{equation}
y_k = \left( \frac{m v_0^2}{a} \right) ^{1/n} x_k, \qquad
\tau = \frac{1}{v_0} \left( \frac{m v_0^2}{a} \right) ^{1/n} t,
\end{equation}
Eq.~(\ref{eq:motion}) can be rewritten as
\begin{eqnarray}
\ddot{x}_k &=& - (x_k - x_{k+1})^{n-1} \theta (x_k -
x_{k+1}) \notag\\
&&+  (x_{k-1} - x_{k})^{n-1} \theta (x_{k-1} -
x_{k}),\label{eq:motion_rescaled}
\end{eqnarray}
where a dot denotes a derivative with respect to $t$.
In the rescaled variables
the initial conditions become $ x_k(0) = \dot{x}_k(0) = 0$, 
$\forall k \neq 1$, $ x_1(0) = 0 $, and $ \dot{x}_1(0) = 1 $. 

When $n>2$ an initial impulse settles into a pulse that
is increasingly narrow with increasing $n$, and
propagates at a velocity that is essentially constant and determined
by $n$ and by the amplitude of the pulse. For $n=2$ the pulse spreads in
time and travels at a constant velocity independent of pulse amplitude.
In the latter case there is considerable backscattering that leads to
backward motion of all the granules behind the pulse, whereas the
backscattering is minimal for $n>2$~\cite{hinch,ourchain}.  The pulse is
a completely conservative solitary wave in the limit $n\rightarrow\infty$.

Three features determine pulse dynamics in these chains:
\begin{enumerate}
\item
The power $n$ in the potential;
\item
The absence of a restoring force; and
\item
The discreteness of the system.
\end{enumerate}
Recently, we discussed the role each of these features in the
continuum approximation and extended previous results to include
viscosity~\cite{ourchain}. Not only does this approximation work very well
for the $n=2$ case~\cite{hinch,ourchain}, but,
corroborating Nesterenko's
theory\cite{nesterenko}, we found that the continuum approximation
works surprisingly well for the prototypical grain, namely spheres.
However, a discussion of the reasons for this agreement seems to be lacking. 
On the other hand, models based on binary interactions have
also been proposed in order to study a chain of spherical and other
grains. Wu's independent-collision model~\cite{wu}
focuses on a chain of tapered grains. From energy and momentum
conservation considerations, working in the $ n\rightarrow \infty $
limit, he shows that his simple model captures the qualitative
behavior observed in simulations for spheres. Subsequently this
model was phenomenologically extended and compared with
experimental results~\cite{nakagawa}. 

\begin{figure}
\vspace*{0.8cm}
\begin{center}
\includegraphics[width=8cm]{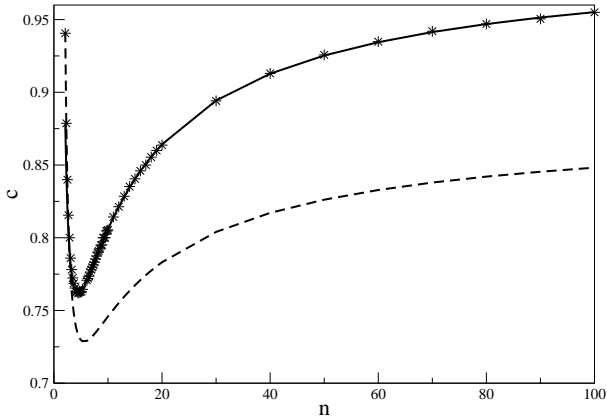}
\caption{Pulse velocity as a function of the power of the
potential. The stars represents the numerical simulation results,
the continuous line the binary collision approximation, and
the broken line the continuum approximation.\label{fig:vel}}
\end{center}
\end{figure} 

Herein we address the question of the applicability of both
the continuum approximation and a binary interaction model
through the analysis of the velocity of the signal propagation
as a function of the power of the potential. For the former case,
the pulse velocity, $C_c(n), $ can be written as~\cite{ourchain}
\begin{equation}
C_c(n) = \sqrt{\frac{2}{n}} \left [ \frac{n^2 (n-2)}{2(n+2)\sqrt{\frac{n(n-1)}{6}} I\left (\frac{4}{n-2}\right )} \right ]^\frac{n-2}{2n},
\label{eq:velcont}
\end{equation} 
\noindent
where
\begin{equation}
I(l)  = 
2^{\,l} \; \frac{\Gamma^2\left(\frac{l+1}{2}\right)}{\Gamma(l+1)}.
\label{eq:int}
\end{equation}

On the other hand, for the binary collision approximation the set
of equations~(\ref{eq:motion_rescaled}) reduces to two coupled equations
which may be decoupled by defining the normal mode variables
$ z_{\pm} = x_1 \pm x_2. $  In particular, we have
\begin{equation}
\ddot{z}_{-} = -2{z}_{-}^{n-1}.
\end{equation} 
This is precisely the equation of motion for one particle
subjected to a potential
$V(z) = 2 z^n/n. $ Furthermore, the initial conditions for the original variables lead to $ z_-(0)=0 $ and $ \dot{z}_-(0)=1. $ Hence, from energy conservation we have:
\begin{equation}
\frac{1}{2}\dot{z}_{-}^2(t) - \frac{1}{2} =  - \frac{2}{n} z_{-}^n(t).
\end{equation} 
Consequently, when the two particles have the same velocity
($\dot{z}_- = 0$), their compression will be a maximum and given 
by $z_m = \left (\frac{n}{4}\right )^{1/n}$. Therefore, the time necessary for the ``pulse'' to travel from the first to the second particle is 
\begin{equation}
T_b(n) = \int_0^{z_m} \frac{1}{\sqrt{1 - 4 z_-^n/n}} \mathrm{d} z_-.
\label{eq:timebinary}
\end{equation} 
Explicitly integrating Eq.~(\ref{eq:timebinary}), we can write
the pulse velocity $C_b(n) = 1/T_b(n), $ as
\begin{equation}
C_b(n) =  \frac{1}{\pi^{1/2}} \left( \frac{4}{n}\right)^{1/n}
\frac{\Gamma(\frac{1}{2}
+\frac{1}{n})}{\Gamma(1+\frac{1}{n})}
\label{eq:velbinary}
\end{equation} 
Our comparison is thus between Eqs.~(\ref{eq:velcont})
and (\ref{eq:velbinary}).

\begin{figure}
\vspace*{0.8cm}
\begin{center}
\includegraphics[width=8cm]{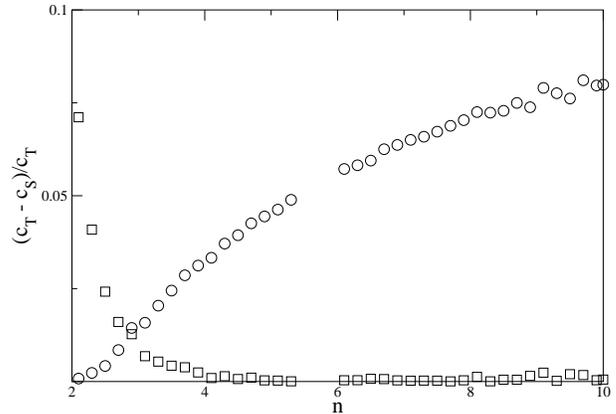}
\caption{Relative error $(C_T-C_S)/C_T$ in the pulse
velocity as a function of the power of the potential. Here, the
theoretical pulse velocity $C_T$, is either the one obtained from
the continuum (circles)
or from the binary collision (squares) approximation. $C_S$ is the pulse
velocity obtained from the numerical simulations of the chain.
\label{fig:vel2}}
\end{center}
\end{figure} 

Qualitatively both approximations give the same result:
for $ n\gtrsim 2 $ the pulse velocity decreases with $n$,
attaining its minimum value for $n \simeq 5$, and then increasing
and saturating
for large $n$ (see Fig.~\ref{fig:vel}). Quantitatively, however,
they differ appreciably. For instance, for
large $n$, $C_b \rightarrow 1$ while $C_c
\rightarrow 0.883\ldots$. In Fig.~\ref{fig:vel2} we present the
relative error of our numerical simulation of the chain as compared
with each theory. From this figure, it is easy to see that while
the continuum approximation gives very good results for small $n$,
its predictions are poor for $n\gtrsim 3.5$. On the other hand,
the binary collision approximation is extremely accurate for
$ n > 3.5$ but not accurate for small $n$.  The quantitative agreement
of each of the two analytic results at their respective $n$ extremes
with the numerical simulations is seen to be excellent.  Again, we
stress that \emph{these results are a reflection of the behavior of the pulse
width}.  At smaller $n$ the pulse is relatively broad, the velocity pulse
covers a number of grains~\cite{ourchain}, and a continuum approximation
captures the pulse configuration and speed very well.  At larger $n$ the
pulse becomes very narrow, discreteness effects dominate the behavior,
and an approximation that assumes that two-grain collisions 
dominate the pulse behavior reproduces the pulse velocity extremely
accurately.

In summary, Nesterenko's continuum approximation gives quantitatively
accurate results for the pulse velocity 
for relatively soft potentials, $n\lesssim 3.5$ (which includes the
generic cases of cylindrical and spherical grains), while the binary
collision model is quantitatively
correct for relatively hard potentials, $n\gtrsim 3.5$.

\section*{Acknowledgments}
This work was supported by the Engineering Research Program of
the Office of Basic Energy Sciences at the U. S. Department of Energy
under Grant No. DE-FG03-86ER13606.

\end{document}